\documentclass[letterpaper]{article} 
\usepackage{tabularx}
\usepackage{amsmath}
\usepackage{multirow}
\usepackage{url}
\usepackage{aaai2026}  
\usepackage{times}  
\usepackage{helvet}  
\usepackage{courier}  
\usepackage{graphicx} 
\usepackage{natbib}  
\usepackage{caption} 
\usepackage{algorithm}
\usepackage{algorithmic}

\usepackage{newfloat}
\usepackage{listings}
\DeclareCaptionStyle{ruled}{labelfont=normalfont,labelsep=colon,strut=off} 
\floatstyle{ruled}
\newfloat{listing}{tb}{lst}{}
\floatname{listing}{Listing}
\title{Membership Inference Attack Against Large Language Model-Based Recommendation Systems: A New Distillation-Based Paradigm}
\author{
    Cuihong Li\textsuperscript{\rm 1}, Xiaowen Huang\textsuperscript{\rm 1,2,3}\thanks{Corresponding Author}, Chuanhuan Yin\textsuperscript{\rm 1}, Jitao Sang\textsuperscript{\rm 1,2,3}\\
}
\affiliations{
    
    \textsuperscript{\rm 1}School of Computer Science and Technology, Beijing Jiaotong University\\
    \textsuperscript{\rm 2}Beijing Key Laboratory of Traffic Data Mining and Embodied Intelligence\\
    \textsuperscript{\rm 3}Key Laboratory of Big Data \& Artificial Intelligence in Transportation, Ministry of Education
}

\usepackage{bibentry}

\begin{document}

\maketitle

\begin{abstract}
Membership Inference Attack (MIA) aims to determine whether a specific data sample was included in the training dataset of a target model. Traditional MIA approaches rely on shadow models to mimic target model behavior, but their effectiveness diminishes for Large Language Model (LLM)-based recommendation systems due to the scale and complexity of training data. This paper introduces a novel knowledge distillation-based MIA paradigm tailored for LLM-based recommendation systems. Our method constructs a reference model via distillation, applying distinct strategies for member and non-member data to enhance discriminative capabilities. The paradigm extracts fused features (e.g., confidence, entropy, loss, and hidden layer vectors) from the reference model to train an attack model, overcoming limitations of individual features. Extensive experiments on extended datasets (Last.FM, MovieLens, Book-Crossing, Delicious) and diverse LLMs (T5, GPT-2, LLaMA3) demonstrate that our approach significantly outperforms shadow model-based MIAs and individual-feature baselines. The results show its practicality for privacy attacks in LLM-driven recommender systems.  
\end{abstract}


\begin{links}
    \link{Code}{https://github.com/Cherie212/MIA4LLMRS.git}
\end{links}

\section{Introduction}

Membership Inference Attack (MIA) is an important type of attack in machine learning. MIA's main objective is to determine whether a specific data sample was used in the training process of a target model \cite{shokri2017membership}. In the context of recommendation systems, an MIA attacker attempts to infer whether a particular user-item interaction record was utilized to train the target recommendation model. MIA attacks in recommendation systems typically follow the shadow model paradigm. The attacker first constructs a shadow model that mimics the behavior of the target model, then uses the shadow model to generate training data for the attack model. Finally, the trained attack model is employed to perform membership inference against the target model.

Large language models (LLMs) have witnessed vigorous development recently, acquiring powerful general capabilities \cite{meta2024llama3} \cite{radford2019language} \cite{raffel2020exploring}. Recommendation systems are progressively evolving towards integration with LLMs. LLM-based recommendation systems refer to recommendation systems that utilize LLMs as core components to improve recommendation performance. Unlike traditional recommendation systems, LLM-based recommendation systems is not limited to ID-based collaborative signals or constrained content features. Instead, it utilize LLM's comprehension capabilities and world knowledge to enhance recommendation performance. A common approach involves fine-tuning LLMs using recommendation datasets to develop LLM-based recommendation systems. 

However, the training and fine-tuning data of LLMs are characterized by enormous volume and diverse sources, which makes the construction of shadow models extremely difficult. Existing work \cite{duan2024membership} has shown that the attack capability of traditional MIA against LLMs is comparable to that of random attacks. LLM-based recommendation systems are essentially LLMs, thus shadow model-based MIA is also less effective against LLM-based recommendation systems. The reference-based MIA \cite{mattern2023membership} against LLMs has good performance, but it needs the training data distribution of the target model \cite{mireshghallah2022quantifying}. This is hardly practical in real-world scenarios. In addition, as far as we know, little work has been done on the MIA against LLM-based recommendation systems.

\begin{figure*}[t]
\centering
\includegraphics[width=0.9\textwidth]{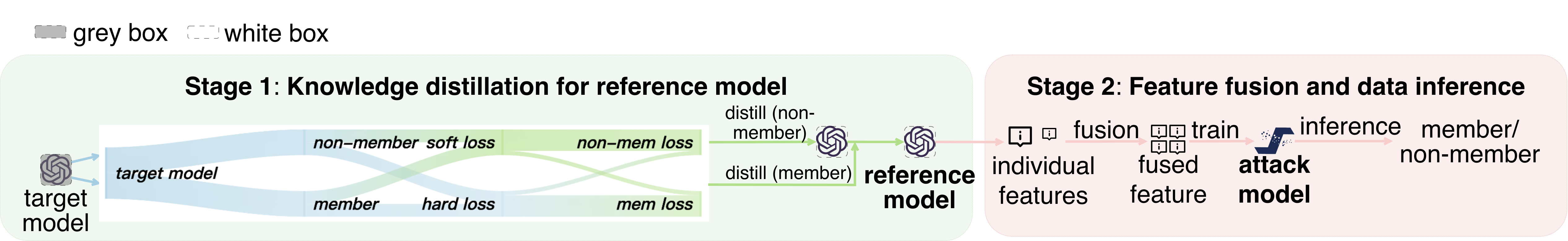} 
\caption{An overview of our paradigm.}
\label{fig1}
\end{figure*}

To address these issues, this paper introduces a novel membership inference attack paradigm for LLM-based recommendation systems. Our paradigm uses knowledge distillation to obtain a reference model, which can generate more discriminative features for member and non-member data compared with traditional shadow model-based methods. The reference model is a model distilled on our background knowledge according to the target model. It is employed to extract features for training the attack model. We apply distinct distillation strategies for members and non-members in the background knowledge respectively. This allows our reference model with enhanced discriminative capability between members and non-members compared to conventional shadow models, because shadow models do not differentiate between member and non-member data during the training process. In addition, we observed that some MIAs against LLMs rely on individual feature to distinguish them, such as calculating the average confidence of partial member data as a threshold to distinguish them \cite{shi2023detecting}. Some other works use different features, such as perplexity \cite{jagannatha2021membership} and sensitivity \cite{maini2024llm}. However, an individual feature is not accurate enough to determine whether the data is included in the training dataset. So we introduce a fused feature to combine multiple features obtained from the reference model to train the attack model, achieving better attack performance than those trained with an individual feature.

In a nutshell, our contributions are as follows:
\begin{itemize}
    \item \textbf{A Novel MIA Paradigm against LLM-based Recommendation Systems:} We propose a new MIA paradigm for LLM-based recommendation systems. It shows significantly enhanced attack effectiveness compared to both shadow model-based MIAs and other LLM-targeted MIAs.
    \item \textbf{Extensive Experiments on Extended Fine-tuning Datasets and Various LLMs: } We expanded the dataset and performed fine-tuning on open-source LLMs of varying scales and architectures, conducting MIA under our paradigm on these fine-tuned models. The detailed results demonstrate the effectiveness of our paradigm in different target models.
    \item Our results demonstrate that the proposed paradigm is both simple and effective for LLM-based recommendation systems. Furthermore, our paradigm shows promising extensibility to other LLM-related tasks.
\end{itemize}

\section{Related Work}

\textbf{Membership Inference Attack.} MIA aims to infer whether a certain data sample was used to train a target model. Traditional MIAs mimicked the behavior of target models by training shadow models \cite{shokri2017membership}, but a recent study has shown that the effectiveness of this paradigm is similar to that of random attacks \cite{duan2024membership}. For MIA against LLMs, some works have focused on data features, using the distributional differences between members and non-member data to distinguish them. Existing works adhere to the principle that member data generally has a lower perplexity (PPL) than non-member data \cite{yeom2018privacy}. Calculating the average PPL of partial member data as a threshold adheres to the principle that the data with a PPL lower than this threshold is classified as member data, and vice versa \cite{carlini2021extracting}. 

\textbf{Fine-tuning for LLM-based Recommendation Systems.} With the rise of LLMs, efforts to combine LLMs with recommendation systems have emerged to leverage large models' rich world knowledge and cross-domain capabilities to enhance recommendation performance \cite{bao2023tallrec} \cite{zhang2024text} \cite{zhang2025collm} \cite{xu2023baize}. Fine-tuning focuses on further training a LLM using task-specific or domain-specific data to get stronger performance. There have been some works on fine-tuning LLM on natural language recommendation datasets to obtain large recommendation models \cite{bao2023tallrec} \cite{zhang2024text} \cite{xu2023baize}.

\textbf{Knowledge Distillation.} Knowledge distillation is typically used to learn a lightweight student model from a large teacher model for model compression \cite{li2021dynamic} \cite{chen2020online} \cite{romero2014fitnets}. A typical response-based distillation model achieves knowledge transfer by calculating the divergence loss of logits between the student model and the teacher model. In addition, soft label-based knowledge distillation methods have also attracted attention \cite{hinton2015distilling}. Soft labels refer to the logits output by the teacher model, while hard labels are the ground truths of the data itself. 

\section{Main Methodology}
As shown in Figure 1, our paradigm is mainly divided into two stages. Stage 1: Knowledge distillation for reference model.  Perform knowledge distillation on the target model using background knowledge to obtain a reference model. Stage 2: Feature fusion and data inference. Obtain features through the reference model, fuse the features, and then training attack model for data inference.

\begin{figure}[t]
\centering
\includegraphics[width=\linewidth]{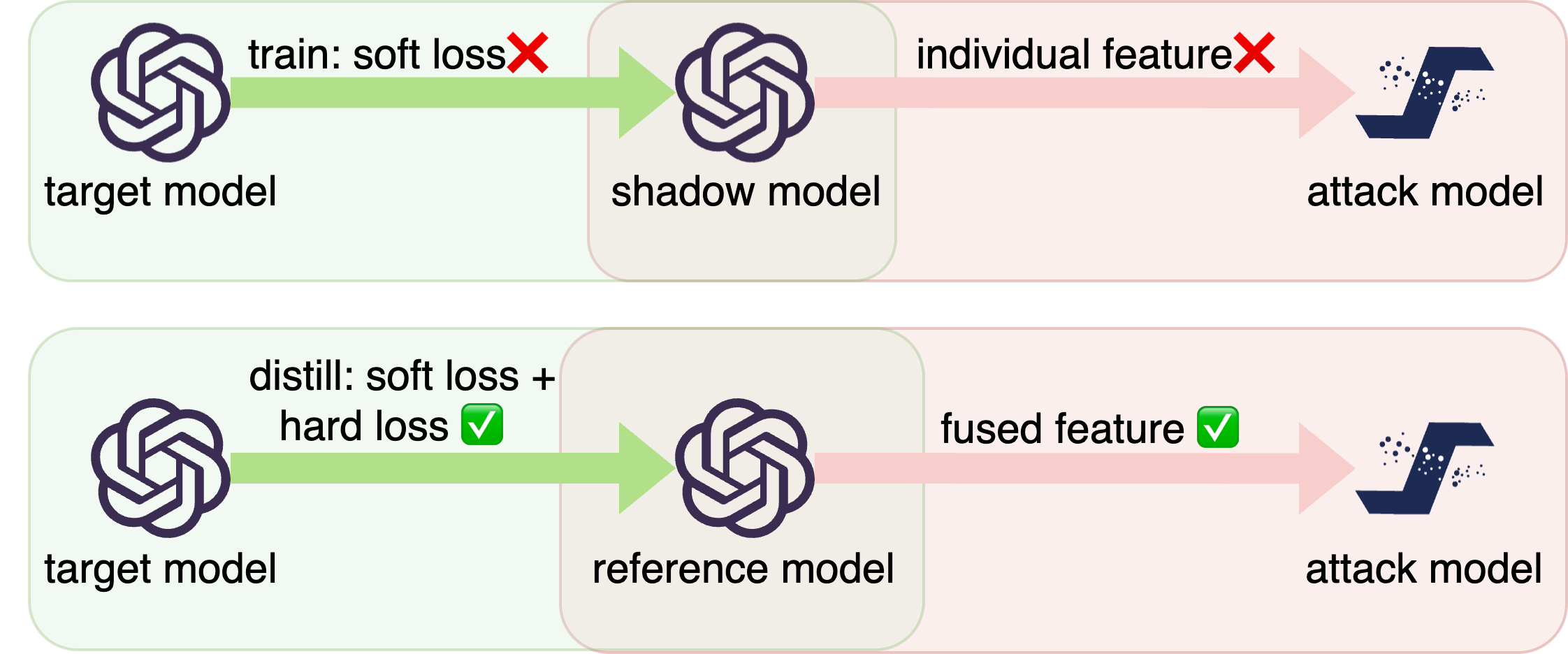}
\caption{The differences between our paradigm (the lower) and shadow model-based MIA (the upper) and the advantages of our approach.}
\label{fig2}
\end{figure}

\subsection{Comparison with Shadow-based MIA}
As shown in Figure 2, the key distinctions between our framework and shadow model-based MIA lie in: (1) the different goal of intermediate models used to generate effective features, and (2) the strategy of soft label and hard label we use and the effective fused features we generate for better attack performance.

From the perspective of intermediate models used to generate effective features, our paradigm generates a more effective reference model compared to shadow model. In shadow model-based MIAs, the shadow model is trained to mimic the target model's behavior, typically by maximizing behavioral consistency between them (e.g., through KL divergence minimization) \cite{shokri2017membership}. However, this paradigm becomes suboptimal when the target model exhibits mediocre performance, since shadow models generally cannot surpass the target model's capabilities. Importantly, in real scenarios, target models include not only well-performing mature LLMs but also those fine-tuned for recommendation tasks, many of which remain underdeveloped in this domain and demonstrate imperfect performance \cite{bao2023tallrec}. 

Our framework fundamentally redefines the reference model's purpose: instead of imitating the target model's behavior, we focus on maximizing the distinction between member and non-member. A white-box reference model that exhibits more pronounced behavioral differences between members and non-members is crucial for extracting effective contrastive features. This is particularly valuable for target models with inherent performance limitations.

From the perspective of data and feature, we use a more effective distillation strategy and generate stronger features. Shadow model-based MIAs mimic the behavior of the target model without distinguishing between member and non-member data during the shadow model training process. And both shadow model-based MIAs and some LLM-targeted MIAs rely on individual feature to train attack models. However, individual feature provides inadequate ability for effectively distinguishing between members and non-members in complex-structured LLMs.

Our framework integrates both soft labels and hard labels with different weights in member and non-member respectively to enhance discriminative capability between member and non-member. We also employ fused features to improve attack performance. Fused feature simultaneously addresses two critical limitations: the poor discriminative performance of an individual feature and suboptimal utilization of different features' discriminative capabilities.

\subsection{Threat Model}
\textbf{Adversary's Goal}. The attacker's objective is to determine whether a given data sample is part of the target model's fine-tuning dataset. A successful MIA must demonstrate effectiveness significantly beyond random attack. 

\textbf{Adversary's Knowledge and Capacity}. In our paradigm, the attacker's background knowledge follows the gray-box attack definition in \cite{wu2025membership}: the attacker masters the model type of the target model (such as natural language models, classification models, etc.), but does not know the architectural details of the model. We hypothesize that the fine-tuned model has assimilated the knowledge from the fine-tuning dataset, making the fine-tuning data and training data equally valuable to an attacker during MIA.

\textbf{Attack Setting}. We employ multiple datasets to simulate real-world attack scenarios. Specifically, for each dataset, we treat 20\% of the whole dataset as the non-member (which is not involved in constructing the target model), while the remaining data serves as the member dataset used to fine-tune the raw model and construct the target model in a recommendation scenario. From the member dataset, we randomly select 5\% of the data to represent the member data known to the attacker, while the remaining member data remains inaccessible to the attacker. We believe this aligns with a common real-world scenario: due to the confidentiality of member data, attackers can only obtain a small portion of it. Meanwhile, they can generally acquire non-member data from public resources by inferring the data format from the limited known member data and observing the target model's outputs.

\subsection{Knowledge Distillation for Reference Model}
For more effective differentiation between member and non-member data, we perform knowledge distillation on member data and non-member data respectively with different strategies. Our reference model employs softmax activation in the output layer and uses cross-entropy loss with one-hot encoded labels for hard label. We employ soft labels and hard labels for each data instance. Hard label and soft label can be formulated as follows:
\begin{equation}
HardLabel_i =
\begin{cases}
[1, 0], & \text{if member.} \\
[0, 1], & \text{if non-member.}
\end{cases}
\end{equation}

\begin{equation}
SoftLabel_i = \frac{\exp ( (logit^{(t)}_i) ^{(T)}/ T )}{\sum_{k=1}^{K} \exp((logit^{(t)}_k) ^{(T)}/ T)},
\end{equation}
where $HardLabel_i$ represents the hard label of the $i$-th data sample, $SoftLabel_i$ represents the soft label of the sample's $i$-th token, $logit^{(t)}_i$ represents the output of the $i$-th logit by the teacher model, $T$ represents the temperature parameter used to smooth the softmax distribution and $K$ represents the vocabulary size of the teacher model.

We compute soft loss and hard loss for both member and non-member data respectively. The soft loss refers to the KL divergence between the logit output by the student model and the teacher model, while the hard loss refers to the cross-entropy loss of the data. Hard loss and soft loss can be formulated as follows:
\begin{equation}
HardLoss_i = - \sum_{j=1}^{2} HardLabel_i \cdot \log (p_j),
\end{equation}
where $HardLoss_i$ represents the hard loss of the $i$-th data instance and $p_j$ represents the predicted probability that the sample belongs to class $j$.

\begin{equation}
SoftLoss_i = T^2 \sum_{i=1}^{K} SoftLabel_i \cdot \Delta_i,
\end{equation}
where $SoftLoss_i$ represents the soft loss of the $i$-th data instance, $T$ is the temperature parameter of distillation and
\begin{equation}
\Delta_i = \log SoftLabel_i - \log (logit^{(s)}_i),
\end{equation}
where $logit^{(s)}_i$ represents the $i$-th logit output of the student model.

For member data, the final loss is computed as:
\begin{equation}
L_{{mem},i} = \alpha \cdot HardLoss_i + (1 - \alpha) \cdot SoftLoss_i ,
\end{equation}
where $\alpha$ controls the relative weight between hard loss and soft loss. This will enhance the performance of the reference model on member data compared to the shadow model, especially when the teacher model does not perform well on member data.

For non-member data, the final loss is computed as: 
\begin{equation}
L_{{nonmem},i} = (1 - \alpha) \cdot HardLoss_i + \alpha \cdot SoftLoss_i,
\end{equation}
which no longer aims to improve the performance of the reference model but imitates the teacher model's behavior, making the distinction between member and non-member data on the reference model more pronounced.

As shown in Algorithm 1, the raw student model is first distilled on non-member data, emphasizing the focus on soft labels to inherit the target model's behavior on non-members regardless of performance quality. This ensures that the reference model's performance on unseen data aligns as closely as possible with the target model. Our goal at this stage is to distinguish between member and non-member data so the poor performance of the reference model on non-member data does not affect our objective.

The obtained model continues to be distilled on member data, emphasizing the focus on hard labels, because the performance on member data needs to be as good as possible to be distinguished from non-members. This mitigates the problem of low discriminability caused by the target model's suboptimal performance. When the reference model achieves strong performance on member data (compared to its near-random performance on non-member data), we can get a distinguishable divergence between member and non-member. The model after the two distillation steps serves as the final reference model for subsequent feature extraction.

\begin{algorithm}[t]
\caption{Knowledge Distillation for Reference Model}
\label{alg:distill}
\begin{algorithmic}[1]
\REQUIRE Teacher model $t$, Student model $s$
\REQUIRE Non-member data $D_{\text{nonmem}} = \{x_1, \dots, x_n\}$  
\REQUIRE Member data $D_{\text{mem}} = \{y_1, \dots, y_m\}$  
\STATE Initialize $\alpha$ and training epochs
\FOR{each epoch} 
    \FOR{ $x_i \in D_{\text{nonmem}}$}
        \STATE Compute $\mathcal SoftLoss_i $, $\mathcal HardLoss_i$
        \STATE Compute $\mathcal L_{{nonmem},i}$
        \STATE Update $s$ with $\mathcal L_{{nonmem},i}$   
    \ENDFOR
\ENDFOR
\FOR{each epoch}
    \FOR{ $y_j \in D_{\text{mem}}$}
        \STATE Compute $\mathcal SoftLoss_j $, $\mathcal HardLoss_j$
        \STATE Compute $\mathcal L_{{mem},j}$
        \STATE Update $s$ with $\mathcal L_{{mem},j}$ 
    \ENDFOR
\ENDFOR
\RETURN $s$
\end{algorithmic}
\end{algorithm}

\subsection{Feature Fusion and Data Inference}

Since the reference model obtained is a white-box model, we can extract any feature from any layer of the model. The penultimate hidden layer generally contains the model's deeper-level, multidimensional abstract features, which generally preserve both semantic and structural information. Therefore, we extract features of member and non-member data through the reference model's penultimate hidden layer and concatenate these features into fused features for training the attack model. We use confidence, entropy, loss, and the logit vector of the penultimate layer as our features to create fused feature.

There exists a significant dimensionality discrepancy among these features: confidence, loss, and entropy are all one-dimensional floating point vectors, whereas the vector feature has 512 dimensions. Direct concatenation would cause the effect of low-dimensional features to be diluted by high-dimensional ones. Therefore, we employ MLP to elevate the dimensions of low-dimensional vectors before concatenating them with the high-dimensional vector feature. We also set different weights for each low-dimensional vectors because they make different contributions to attack performance. The fused features are used to train attack model. After being trained, the attack model is used for final inference.

\section{Experiments}
\subsection{Experimental Setup}
\textbf{Target Models.} From a model architecture perspective, we aim to select a comprehensive set of models for our experiments. We choose the T5 \cite{raffel2020exploring} from the encoder-decoder architectures for experimentation. Among mainstream decoder-only architectures, we choose GPT-2 \cite{radford2019language} and LLaMA3 \cite{meta2024llama3}. Our target model is fine-tuned using LoRA (Low-Rank Adaptation), a parameter-efficient fine-tuning method that enables LLM to acquire specific knowledge from the fine-tuning dataset \cite{hu2022lora}.

\begin{table}[h!]
\centering
\setlength{\tabcolsep}{2.5pt} 
\begin{tabular}{c|c|c|c|c}
\hline &
\multicolumn{1}{c|}{Last.FM} &
\multicolumn{1}{c|}{Movie} &
\multicolumn{1}{c|}{Book} &
\multicolumn{1}{c}{Delicious} \\
\hline
\#original & 2k+ & 8k & 19k+  &114k+\\
\#non-member & 1k+ & 2k & 4k+  &76k+\\
\#member & 100 & 400 & 900  &5k+\\
\hline
\end{tabular}
\caption{Dataset illustration. ``Movie'' refers to MovieLens and ``Book'' refers to Book-Crossing.}
\label{table1}
\end{table}

\begin{table*}[h!]
\centering
\setlength{\tabcolsep}{2.5pt}  
\begin{tabular}{c|cccc|cccc|cccc|cccc}
\hline
\multicolumn{1}{c|}{\multirow{2}{*}{}} &
\multicolumn{4}{c|}{Last.FM} &
\multicolumn{4}{c|}{MovieLens} &
\multicolumn{4}{c|}{Book-Crossing} &
\multicolumn{4}{c}{Delicious} \\
\cline{2-17}
& acc\% & recall\% & f1\% & auc\% & acc\% & recall\% & f1\% & auc\% & acc\% & recall\% & f1\% & auc\% & acc\% & recall\% & f1\% & auc\% \\
\hline
P & 59.76 & 54.55 & 11.24 & 52.41 & 62.33 & 74.50 & 28.42 & 48.94 & 42.96 & 60.93 & 26.24 & 49.41 & 49.54 & 71.57 & 12.64 & 50.25 \\
m & 62.85 & 27.27 & 9.23 & 52.79 & 51.46 & 44.25 & 23.30 & 49.90 & 65.64 & 28.14 & 21.44 & 49.37 & 48.05 & 52.03 & 12.26 & 50.30 \\
m+ & 42.27 & 50.91 & 46.86 & 65.46 & 26.87 & 39.75 & 35.22 & 85.98 & 28.56 & 44.54 & 38.40 & 87.05 & 39.64 & 6.72 & 10.02 & 62.73 \\
Z & 40.37 & 54.54 & 11.25 & 45.95 & 53.21 & 49.00 & 25.87 & 51.90 & 51.25 & 49.79 & 25.39 & 46.69 & 44.64 & 55.49 & 12.27 & 49.33 \\
S & 59.09 & 63.64 & 60.87 & 63.84 & 60.00 & 55.00 & 57.89 & 67.74 & 53.86 & 55.15 & 54.45 & 55.36 & 51.35 & 51.96 & 51.64 & 51.79 \\
O & \textbf{65.09} & \textbf{77.27} & \textbf{73.68} & \textbf{66.74} & \textbf{85.00} & \textbf{93.75} & \textbf{86.21} & \textbf{90.14} & \textbf{86.86} & \textbf{98.97} & \textbf{88.28} & \textbf{93.59} & \textbf{66.36} & \textbf{86.63} & \textbf{72.71} & \textbf{68.20} \\
\hline
\end{tabular}
\caption{Comparative results of our method with baselines across all datasets using T5-base as the target model. ``P'', ``m'', ``m+'', ``Z'', ``S'' and ``O'' mean PPL, min-k\%, min-k\%++, Zlib, Shadow and Ours respectively. The best results for each metric in each dataset are in \textbf{bold}.}
\label{table2}
\end{table*}

\begin{figure*}[h!]
\centering
\includegraphics[width=0.9\textwidth ]{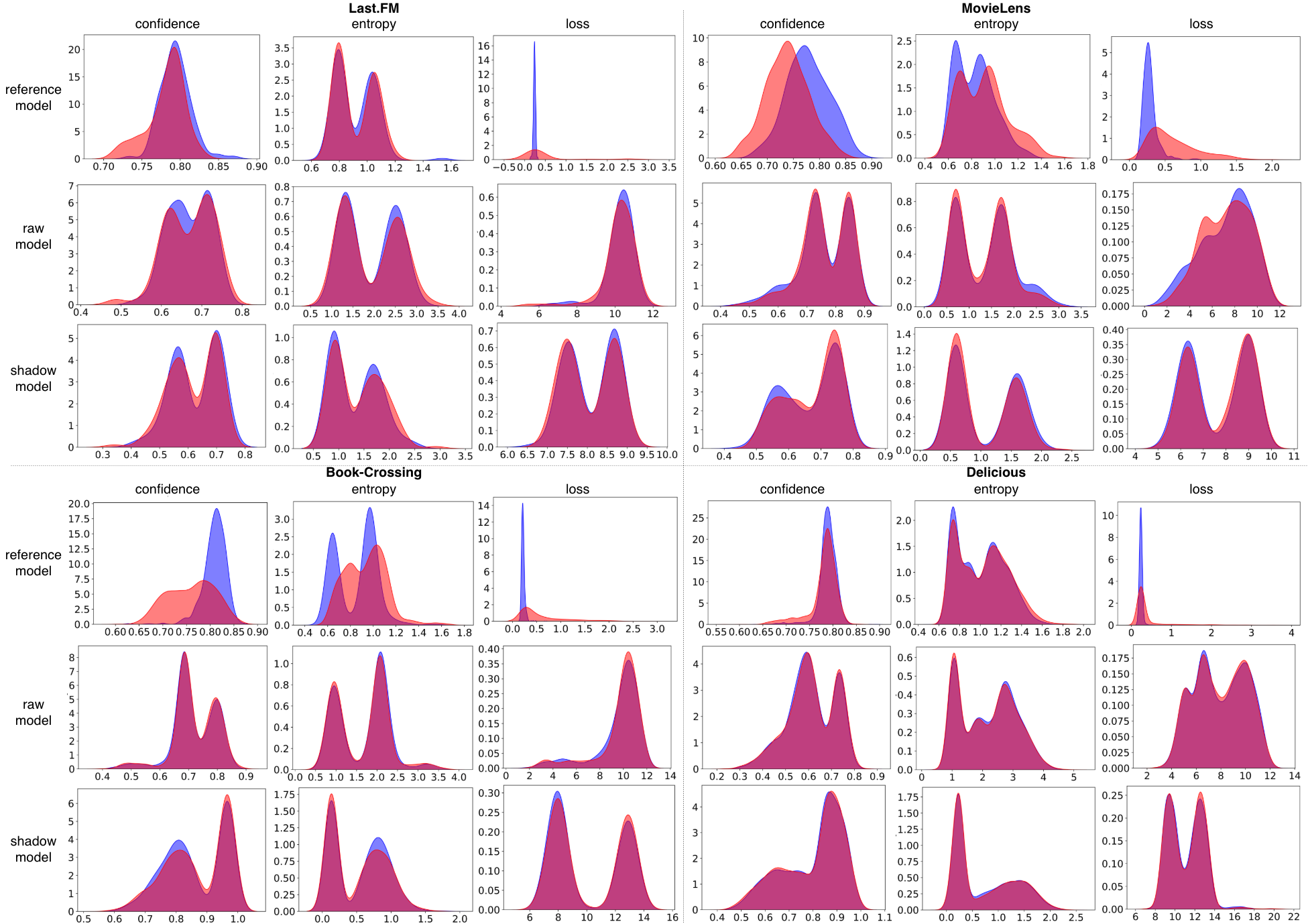}
\caption{The feature distributions of the reference model (T5) in our paradigm, the model before distillation (raw model) and the shadow model. The blue and red areas in each part represent the distribution of members and non-members respectively. The horizontal axis of each picture represents the value of the feature, and the vertical axis represents the data density corresponding to the feature.}
\label{fig3}
\end{figure*}

\textbf{Datasets.} Some works have transformed traditional recommendation data formats (MovieLens and Book-Crossing datasets) into natural language formats to facilitate LLM fine-tuning \cite{bao2023tallrec} \cite{zhang2024text} \cite{zhang2025collm}. Based on this, we convert the Delicious \cite{delicious_web} and Last.FM \cite{lastfm2023} datasets into natural language instruction datasets. We set 5\% of the fine-tuning data as our background knowledge (member) and we have data from public resources (non-member). The original sizes of each dataset and the dataset sizes of the background knowledge (member and non-member) we have are shown in Table 1.

\textbf{Baselines.} We choose the Loss Attack \cite{yeom2018privacy} as one of our baselines, and in the context of LLM, the loss corresponds to the sample's perplexity (PPL). We also choose the min-k\% \cite{shi2023detecting}, which judges members and non-members based on text likelihood token probabilities, and its normalized variant min-k\%++ \cite{zhang2024min} that normalizes text likelihood. We choose the membership inference method \cite{carlini2021extracting}, which compares sample perplexity with Zlib compression entropy. We also compare the performance of our paradigm with that of the shadow model-based MIA.

\textbf{Evaluation Protocols.} We follow the previous studies, using accuracy, recall, f1 score and auc as our evaluation metrics \cite{yeom2018privacy} \cite{shi2023detecting} \cite{zhang2024min}.

\textbf{Implementation details.} During knowledge distillation, each student model undergoes 5 training epochs on member and non-member data respectively. During training the attack model, we randomly select data from non-member data in an amount equal to that of member data for attack model training to avoid data imbalance. We use a logistic regression classifier as our attack model and set the maximum number of iterations to 1000. We employ two linear layers and one ReLU activation function for constructing a fused feature. All of our experiments are conducted on eight A3090 GPUs.

\subsection{Main Results}
We firstly compare the performance of our method and baselines using the T5 model on each dataset. As is shown in Tables 2, our method outperforms the baselines in all cases. In the recommendation field, LLM's outputs are often simple in many cases. For example, user preference judgments may only yield outputs like ``Yes.'' or ``No.''. When the output length is short, the mean and variance of the min-k\%++ method can't accurately reflect the feature distribution, leading to significant discrepancies across metrics. 

The attack performance of shadow model-based MIA against LLM is basically similar to that of random attacks, which also proves that the MIA paradigm based on shadow models is no longer suitable for LLMs \cite{duan2024membership}. 

Other baselines generally rely on individual-feature attacks, which fail to effectively distinguish between members and non-members when the textual similarity between samples is high. For instance, when two data samples serve as member and non-member data respectively, they may differ by only one item in the user's preferences and history while being identical in all other items and descriptions. In such scenarios, our fused feature demonstrates significant advantage over individual features.

To verify the effectiveness of knowledge distillation, we tested the feature distribution (confidence, entropy, and loss) of output for members and non-members on the shadow model, the raw model (before knowledge distillation), and our reference model (after knowledge distillation). The density is derived from kernel density estimation (KDE). As shown in Figure 3, the feature distributions on the shadow model and the raw model exhibit little nuance, indicating that these two models have a low degree of distinction between members and non-members. In contrast, our reference model shows relatively differences in the feature distributions of member and non-member data outputs, which demonstrates the effectiveness of knowledge distillation. We also found that T5's member and non-member output distributions tended toward multimodal patterns in some cases, whereas other models universally demonstrated unimodal Gaussian distributions. While this does not affect the validity of our conclusion about the effectiveness of the reference model, it constitutes an issue worthy of further investigation. A possible reason may be architectural differences.

\subsection{Analytical Study}

\textbf{Demonstration of the Necessity of Our Paradigm}.
We conducted membership inference attacks on several common LLMs (T5, GPT-2 and LLaMA3) using some baselines from our paper, with the results shown in Tables 3 and 4. Most attack metrics are comparable to random attacks (50\%), and some even perform significantly worse, such as the F1 score when attacking the T5 model and the AUC when attacking GPT-2 and LLaMA3. Attack methods that show high performance on certain metrics cannot maintain strong attack effectiveness across all remaining metrics. These findings reveal that none of the current popular MIA methods consistently excels across all metrics when applied to LLMs. This demonstrates the necessity of our work, which aims to establish a new membership inference attack paradigm specifically designed for target models that are essentially LLMs.

\begin{table}[t]
\centering
\setlength{\tabcolsep}{2.5pt} 
\label{table3}
\begin{tabular}{c|c|ccccc}
\hline
Model & Method & acc\% & recall\% & f1\% & auc\% \\
\hline
\multirow{3}{*}{T5} 
& PPL & \textbf{62.33} & \textbf{74.50} & 28.42 & 48.94 \\
& min-k\% & 51.46 & 44.25 & 23.30 & 49.90 \\
& min-k\%++ & 26.87 & 39.75 & \textbf{35.22} & \textbf{85.98} \\
\hline
\multirow{3}{*}{GPT-2} 
& PPL & \textbf{56.33} & 51.50 & 28.22 & 39.31 \\
& min-k\% & 41.71 & \textbf{80.25} & 31.46 & 39.29 \\
& min-k\%++ & 52.50 & 51.90 & \textbf{51.25} & \textbf{48.12} \\
\hline
\multirow{3}{*}{LLaMA3} 
& PPL & \textbf{91.71} & 50.25 & \textbf{66.89} & 34.20 \\
& min-k\% & 80.04 & \textbf{90.04} & 64.99 & \textbf{49.32} \\
& min-k\%++ & 74.50 & 67.72 & 53.50 & 9.13 \\
\hline
\end{tabular}
\caption{Comparison of the attack performance between different baselines on MovieLens dataset across various models.}
\end{table}

\begin{table}[t]
\centering
\setlength{\tabcolsep}{2.5pt} 
\label{table4}
\begin{tabular}{c|c|ccccc}
\hline
Model & Method & acc\% & recall\% & f1\% & auc\% \\
\hline
\multirow{3}{*}{T5} 
& PPL & 42.96 & \textbf{60.93} & 26.24 & 49.41 \\ 
& min-k\% & \textbf{65.64} & 28.14 & 21.44 & 49.37 \\
& min-k\%++ & 28.56 & 44.54 & \textbf{38.40} & \textbf{87.05} \\
\hline
\multirow{3}{*}{GPT-2} 
& PPL & 47.80 & \textbf{58.45} & 27.17 & 47.02 \\ 
& min-k\% & \textbf{51.49} & 53.61 & 26.91 & 47.02 \\ 
& min-k\%++ & \textbf{51.49} & 52.21 & \textbf{52.99} & \textbf{49.09} \\
\hline
\multirow{3}{*}{LLaMA3} 
& PPL & \textbf{69.78} & 55.77 & 38.07 & 27.61 \\ 
& min-k\% & 57.61 & \textbf{81.24} & 38.96 & 27.64 \\ 
& min-k\%++ & 53.61 & 53.37 & \textbf{53.09} & \textbf{46.88} \\
\hline
\end{tabular}
\caption{Comparison of the attack performance between different baselines on Book-Crossing dataset across various models.}
\end{table}

\begin{table*}[h!]
\centering
\setlength{\tabcolsep}{2.5pt} 
\begin{tabular}{c|c|cccc|cccc|cccc}
\hline
\multirow{2}{*}{Data} & \multirow{2}{*}{Student} & \multicolumn{4}{c|}{T5} & \multicolumn{4}{c|}{GPT-2} & \multicolumn{4}{c}{LLaMA3} \\
\cline{3-14}
 & & acc\% & recall\% & f1\% & auc\% & acc\% & recall\% & f1\% & auc\% & acc\% & recall\% & f1\% & auc\% \\
\hline
\multirow{3}{*}{LF} 
& T5 & \textbf{57.35} & \textbf{71.32} & \textbf{63.38} & \textbf{60.22} & 43.18 & 45.45 & 44.44 & \textbf{47.52} & 36.36 & 31.82 & 33.33 & 33.88 \\
& GPT-2 & 50.00 & 45.45 & 47.62 & 39.88 & \textbf{52.27} & \textbf{46.15} & \textbf{53.93} & 34.50 & 45.45 & 50.00 & 47.83 & 51.03 \\
& LLaMA3 & 34.09 & 36.36 & 35.56 & 39.05 & 43.18 & 40.91 & 41.86 & 43.39 & \textbf{63.64} & \textbf{77.27} & \textbf{68.00} & \textbf{65.70} \\
\hline
\multirow{3}{*}{ML} 
& T5 & \textbf{83.13} & \textbf{93.17} & \textbf{84.75} & \textbf{90.42} & 56.25 & 55.00 & 55.70 & 61.62 & 59.38 & 58.75 & 59.12 & 59.50 \\
& GPT-2 & 60.62 & 55.00 & 58.28 & 63.38 & \textbf{67.50} & \textbf{73.75} & \textbf{69.41} & 68.66 & 68.75 & 66.25 & 67.95 & 73.02 \\
& LLaMA3 & 45.62 & 52.50 & 49.12 & 49.59 & 66.87 & 68.75 & 67.48 & \textbf{75.59} & \textbf{94.37} & \textbf{93.75} & \textbf{94.34} & \textbf{99.05} \\
\hline
\multirow{3}{*}{BC} 
& T5 & \textbf{84.28} & \textbf{93.81} & \textbf{85.65} & \textbf{89.58} & 47.16 & 47.94 & 47.57 & 47.03 & 47.68 & 42.27 & 44.69 & 47.45 \\
& GPT-2 & 50.52 & 48.45 & 49.47 & 51.92 & 49.23 & \textbf{52.58} & \textbf{50.87} & \textbf{50.85} & 50.77 & 50.00 & 50.39 & 48.96 \\
& LLaMA3 & 51.03 & 53.61 & 52.26 & 48.99 & \textbf{51.03} & 48.97 & 50.00 & 49.31 & \textbf{69.59} & \textbf{74.23} & \textbf{70.94} & \textbf{77.46} \\
\hline
\multirow{3}{*}{DL} 
& T5 & \textbf{66.36} & \textbf{86.63} & \textbf{72.71} & \textbf{68.20} & \textbf{49.65} & 45.00 & 48.05 & 48.99 & 48.56 & 43.65 & 46.76 & 48.39 \\
& GPT-2 & 49.87 & 48.86 & 50.22 & 49.87 & 49.48 & 48.19 & \textbf{49.67} & \textbf{49.28} & 50.48 & 48.28 & 50.22 & 51.28 \\
& LLaMA3 & 48.74 & 39.78 & 44.54 & 47.84 & 47.39 & \textbf{48.61} & 48.88 & 46.93 & \textbf{57.40} & \textbf{56.43} & \textbf{57.82} & \textbf{60.05} \\
\hline
\end{tabular}
\caption{Comparative results of different types of teacher models and student models on each dataset. The first horizontal and the second vertical axes represent teacher models and student models respectively. ``LF'', ``ML'', ``BC'' and ``DL'' mean Last.FM, MovieLens, Book-Crossing and Delicious respectively. The best results under each teacher model on each dataset are in \textbf{bold}.}
\label{table5}
\end{table*}

\textbf{Distillation Model Type Selection.} We analyzed the impact of different model architectures on distillation. For teacher models, we selected T5-base, GPT-2-medium and LLaMA3-3B, and for student models, we selected T5-small, GPT-2, and LLaMA3-1B. When using each teacher model, we employ different student models. When distillation is conducted between architecturally heterogeneous models, vocabulary mismatch often leads to distillation failure. To address this, we propose vocabulary truncation that prunes the larger vocabulary to match the smaller target vocabulary.

As shown in Table 5, in most cases, student models with architectures similar to their teacher models achieve better attack performance, because vocabulary truncation across differently sized vocabularies inevitably incurs information loss during distillation, significantly degrading knowledge transfer efficiency.

\textbf{Effect of Knowledge Distillation Parameter}. We analyzed the impact of $\alpha$ during the knowledge distillation. We varied $\alpha$ from 0 to 1 in increments of 0.1. When $\alpha$ is between 0.1 and 0.9, we combine both hard loss and soft loss with different weights for each data instance. When $\alpha$ is set to 0 or 1, distillation for a specific data instance uses only hard loss or soft loss. 

As shown in Figure 4, we found that the impact of different $\alpha$ values on the attack performance does not show a completely consistent trend across datasets. For instance, on Last.FM, the attack effectiveness fluctuates significantly with varying $\alpha$ values, and on MovieLens, the optimal attack performance is achieved when $\alpha$ increases to 0.9. For Book-Crossing and Delicious, the best results occur when $\alpha$ equals 1. Therefore, in practical applications, selecting an appropriate $\alpha$ value according to the specific dataset can achieve optimal attack performance. In fact, the determination of $\alpha$ is influenced by various factors, including the model architecture and dataset characteristics, as the implementation focus varies depending on the specific data and task. Although the optimal values of $\alpha$ vary across different datasets, by changing it, our paradigm can achieve superior attack performance compared to a random attack in the four datasets.

\begin{figure}[h!]
\centering
\includegraphics[width=\linewidth]{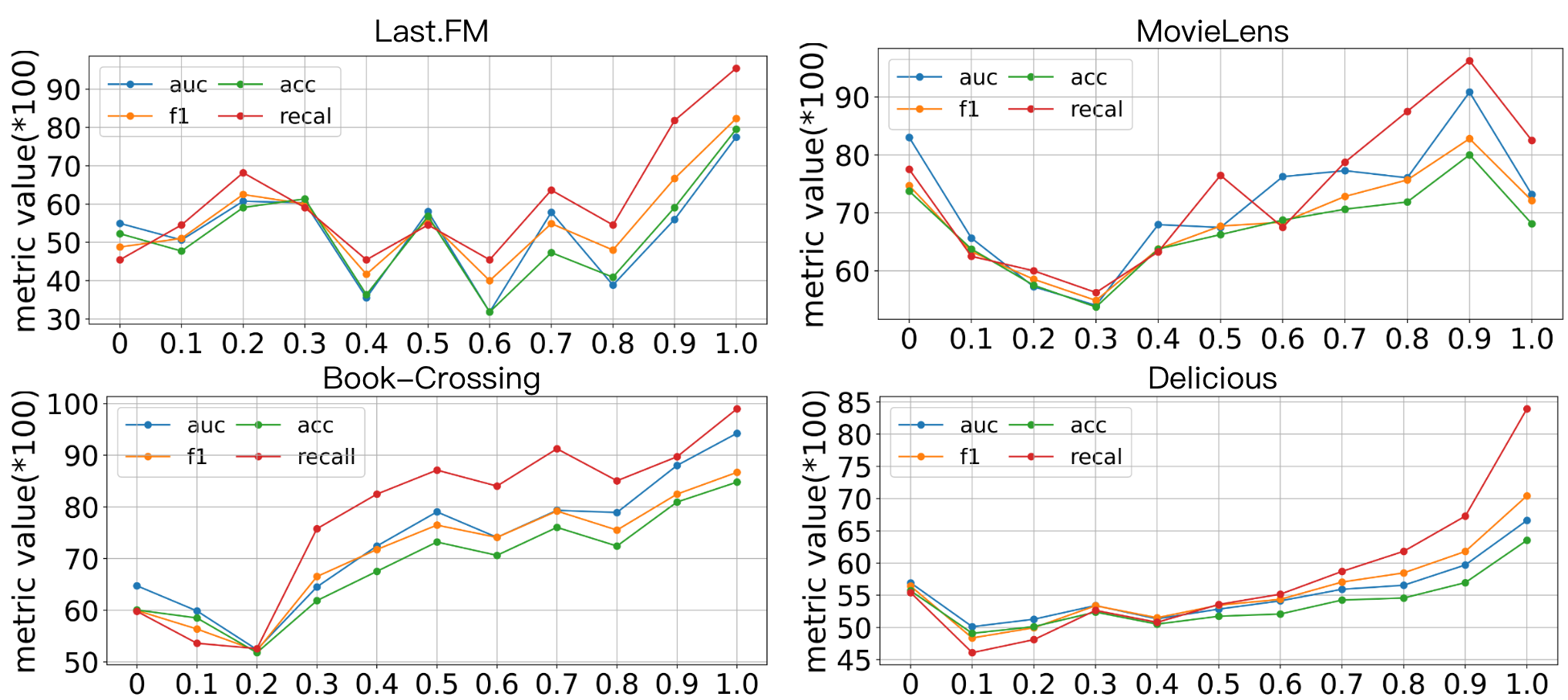}
\caption{The impact of different $\alpha$ values on the performance of our paradigm on the four datasets respectively. The values on the horizontal axis represent the range of $\alpha$ values.}
\label{fig4}
\end{figure}

\begin{table}[t]
\centering
\setlength{\tabcolsep}{1.5pt} 
\begin{tabular}{c|cc|cc|cc|cc}
\hline
\multirow{2}{*}{Ms} & 
\multicolumn{2}{c|}{LF} & 
\multicolumn{2}{c|}{ML} &
\multicolumn{2}{c|}{BC} & 
\multicolumn{2}{c}{DL} \\
\cline{2-9}
 & logist & bert & logist & bert & logist & bert & logist & bert \\
\hline
ac$\uparrow$ & \textbf{65.09} & 59.09 & \textbf{85.00} & 70.00 & 86.86 & \textbf{89.43} & 65.27 & \textbf{66.71} \\
re$\uparrow$ & 77.27 & \textbf{90.91} & \textbf{93.75} & 84.14 & \textbf{98.97} & 92.71 & 87.72 & \textbf{88.35} \\
f1$\uparrow$ & \textbf{73.68} & 68.97 & \textbf{86.21} & 74.19 & 88.28 & \textbf{89.64} & 72.33 & \textbf{72.95} \\
au$\uparrow$ & 66.74 & \textbf{69.63} & \textbf{90.14} & 75.04 & 93.59 & \textbf{93.92} & 68.45 & \textbf{69.53} \\
\hline
tr$\downarrow$ & \textbf{0.12} & 39.08 & \textbf{0.44} & 71.57 & \textbf{1.89} & 143.61 & \textbf{16.17} & 691.36 \\
ev$\downarrow$ & \textbf{0.02} & 0.38 & \textbf{0.06} & 1.77 & \textbf{0.15} & 1.49 & \textbf{0.65} & 11.82 \\
\hline
\end{tabular}
\caption{Comparative performance, training time and evaluation time of different types of attack models on all datasets. ``Ms'', ``ac'', ``re'', ``f1'' and ``au'' represent ``Metrics'', ``acc\%'', ``recall\%'', ``f1\%'' and ``auc\%'' respectively. ``tr'' and ``ev'' represent the attack model's training time (s) and evaluation time (s) respectively. The best results on each dataset are in \textbf{bold}.}
\label{table6}
\end{table}

\begin{table}[h!]
\centering
\setlength{\tabcolsep}{3.5pt} 
\label{table7}
\begin{tabular}{c|c|ccccc}
\hline
Data & Metrics & Ours & Vec & Entr & Conf & Loss \\
\hline
\multirow{4}{*}{LF} 
& acc\% & \textbf{79.55} & 50.00 & 45.45 & 65.91 & 52.27 \\
& recall\% & \textbf{95.45} & 45.45 & 40.91 & 72.73 & 95.00 \\
& f1\% & \textbf{82.35} & 47.62 & 42.86 & 68.09 & 66.67 \\
& auc\% & \textbf{77.48} & 46.90 & 50.00 & 70.87 & 72.73 \\
\hline
\multirow{4}{*}{ML} 
& acc\% & \textbf{80.00} & 59.38 & 58.75 & 69.37 & 78.75 \\
& recall\% & \textbf{96.25} & 62.50 & 71.25 & 75.00 & 91.25 \\
& f1\% & \textbf{82.80} & 60.61 & 63.33 & 71.01 & 81.11 \\
& auc\% & \textbf{90.84} & 61.66 & 62.91 & 74.19 & 84.36 \\
\hline
\multirow{4}{*}{BC} 
& acc\% & \textbf{86.60} & 51.29 & 84.54 & 81.19 & 85.05 \\
& recall\% & \textbf{98.99} & 52.58 & 95.88 & 98.45 & 98.97 \\
& f1\% & \textbf{88.02} & 51.91 & 86.11 & 83.96 & 86.88 \\
& auc\% & \textbf{96.71} & 53.71 & 93.54 & 94.14 & 95.34 \\
\hline
\multirow{4}{*}{DL} 
& acc\% & \textbf{64.75} & 55.48 & 61.97 & 60.31 & 63.84 \\
& recall\% & 88.81 & 52.14 & 89.99 & 93.10 & \textbf{95.04} \\
& f1\% & 72.28 & 54.79 & 71.00 & 70.83 & \textbf{73.12} \\
& auc\% & \textbf{67.55} & 56.79 & 63.99 & 62.69 & 66.08 \\
\hline
\end{tabular}
\caption{Comparison of the attack performance between our fused features and individual features on the four datasets. ``Vec'', ``Entr'' and ``Conf'' mean Vector, Entropy and Confidence respectively.}
\end{table}
\textbf{Attack Model Type Selection.} We analyzed the impact of the complexity of the attack model on attack perfromance. We selected a simple logistic classifier and a relatively complex model architecture for training the attack model. To be precise, we choose BERT \cite{devlin2019bert} to investigate the impact of attack model complexity on both attack effectiveness and time cost. 

As shown in Table 6, when the data volume is small, logistic classifier can meet classification requirements with better performance than BERT. This is because complex models tend to underfit when the dataset is small, leading to poor training performance. As the data volume increases, complex models can marginally improve attack performance, but the improvement is limited while training and inference time costs increase significantly. For instance, when the dataset size exceeds 10k samples (Delicious dataset), BERT's training time is over 40 times longer than that of a logistic regression classifier and its evaluation time is 18 times longer. Considering the little performance improvement and the cost, simple classifiers are sufficient to meet attack requirements, which also validates the significance of the distilled models and features.

\textbf{Effect of Fused Features.} To evaluate the effectiveness of fused features, we compared the attack performance of our attack model with those trained using individual features (Table 7) and those trained after removing each feature (Table 8). In Table 7, our method outperforms attack models trained with individual features in most situations, indicating that fused feature can accumulate the differences in individual features between member and non-member data. In Table 8, our method also surpasses most attack models trained after removing any individual feature, indicating that each feature in the fused feature contributes to improving the attack performance.

The results also demonstrate that different features contribute variably to the attack effectiveness. Among the four features we employed, loss exhibits the highest contribution to the attack performance, whereas vector shows the minimal enhancement effect.

\begin{table}[t]
\centering
\setlength{\tabcolsep}{3.5pt} 
\label{table8}
\begin{tabular}{c|c|ccccc}
\hline
Data & Metrics & Ours & /Vec & /Entr & /Conf & /Loss \\
\hline
\multirow{4}{*}{LF} 
& acc\% & \textbf{79.55} & 70.45 & 72.73 & 59.09 & 54.55 \\
& recall\% & \textbf{95.45} & 95.45 & 90.91 & 81.82 & 50.00 \\
& f1\% & \textbf{82.35} & 76.36 & 76.92 & 66.67 & 52.38 \\
& auc\% & \textbf{77.48} & 63.22 & 65.91 & 58.88 & 59.30 \\
\hline
\multirow{4}{*}{BC} 
& acc\% & \textbf{86.60} & 85.31 & 84.79 & 84.28 & 85.57 \\
& recall\% & \textbf{98.99} & 98.45 & 98.45 & 97.94 & 97.42 \\
& f1\% & \textbf{88.02} & 87.02 & 86.62 & 86.17 & 87.10 \\
& auc\% & \textbf{96.71} & 94.28 & 91.48 & 92.86 & 93.08 \\
\hline
\multirow{4}{*}{ML} 
& acc\% & 80.00 & \textbf{80.63} & 79.37 & 80.00 & 69.37 \\
& recall\% & \textbf{96.25} & 92.50 & 90.00 & 91.25 & 75.00 \\
& f1\% & \textbf{82.80} & 82.68 & 81.36 & 82.02 & 71.01 \\
& auc\% & \textbf{90.84} & 86.36 & 85.95 & 85.27 & 75.42 \\
\hline
\multirow{4}{*}{DL} 
& acc\% & \textbf{64.75} & 64.23 & 63.01 & 63.49 & 62.27 \\
& recall\% & \textbf{88.81} & 86.80 & 85.87 & 86.12 & 80.32 \\
& f1\% & \textbf{72.28} & 71.52 & 70.61 & 70.94 & 68.78 \\
& auc\% & 67.55 & \textbf{68.29} & 66.03 & 66.83 & 64.48 \\
\hline
\end{tabular}
\caption{Comparison of the performance between our fused feature and that after removing each feature respectively on the four datasets. ``/*'' means removing ``*'' from the fused feature.}
\end{table}

\begin{table*}[t]
\centering
\setlength{\tabcolsep}{2.5pt} 
\begin{tabular}{c|cc|cc|cc|cc}
\hline
 & Con & Wei\_Con & Sum & Wei\_Sum & MLP & Wei\_MLP & MLP\_Com & Wei\_MLP\_Com \\
\hline
acc\% & 80.63 & \textbf{86.00} & 76.88 & \textbf{80.00} & 83.75 & \textbf{85.00} & 76.25 & \textbf{78.75} \\
recall\% & 92.50 & \textbf{95.00} & 90.00 & \textbf{97.50} & 91.25 & \textbf{93.75} & 91.02 & \textbf{93.20} \\
f1\% & 82.68 & \textbf{87.86} & 79.56 & \textbf{82.98} & 84.88 & \textbf{86.21} & 80.22 & \textbf{81.52} \\
auc\% & 89.89 & \textbf{90.87} & 80.95 & \textbf{87.75} & 91.38 & \textbf{90.14} & \textbf{88.06} & 87.73 \\
\hline
\end{tabular}
\caption{Comparison of the performance of different feature fusion strategies on MovieLens. ``Con'' means ``Concatenation'', ``Wei'' means ``Weighted'' and ``Com'' means ``Compression''. The best results in weight inclusion comparisons are in \textbf{bold}.}
\label{table9}
\end{table*}

\begin{figure*}[t]
\centering
\includegraphics[width=0.9\textwidth]{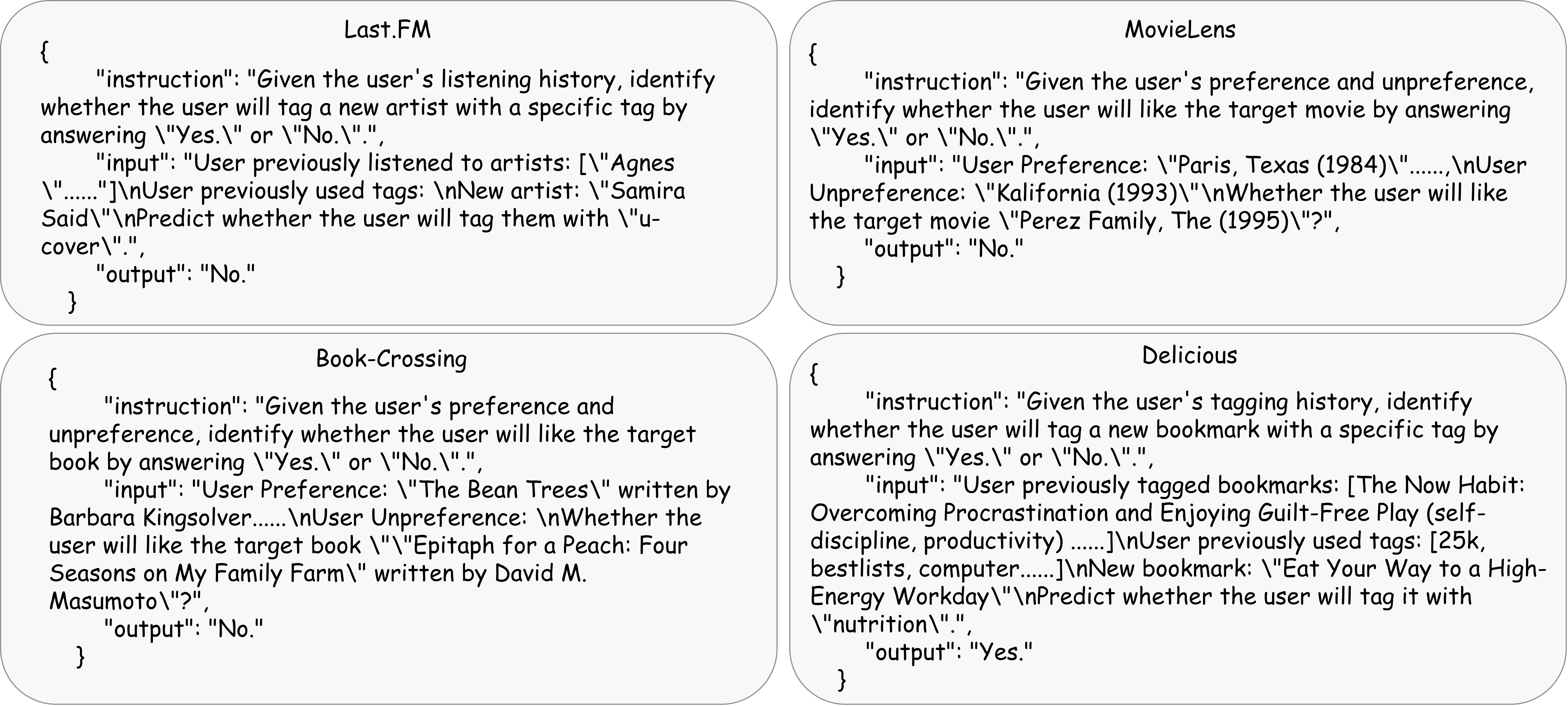}
\caption{Data sample demonstration of the four datasets.}
\label{fig5}
\end{figure*}

\textbf{Feature Fusion Strategy.} 
We analyzed the impact of different feature fusion strategies on attack performance. Our analysis considers the influence of feature weight and dimension on the attack performance. In particular, we consider the dimensional differences in the features we use. The fusion methods that we adopted include: 
\begin{itemize}
    \item \textbf{Concatenation}: Directly concatenating the four features. 
    \item \textbf{Weighted Concatenation}: Different weights are assigned to confidence, entropy, and loss respectively, and then concatenate these features to form a fused feature.
    \item \textbf{Sum}: Add confidence, entropy and loss as a feature and then concatenate it with the vector feature. 
    \item \textbf{Weighted Sum}: Based on Sum, different weights are assigned to confidence, entropy, and loss respectively before adding them into one feature.
    \item \textbf{MLP}: Confidence, entropy and loss (floating-point features) are elevated respectively to the same dimensionality as vector (512D), then concatenating the four 512D vectors to obtain a 4*512D vector. All MLPs use two linear layers and one ReLU activation function.
    \item \textbf{Weighted MLP}: Based on the MLP method above, we perform weighted concatenation of the four vectors to obtain a 4*512D weighted vector. 
    \item \textbf{MLP Compression}: Based on the MLP method, we remap the obtained 4*512D vectors back to 512 dimensions to avoid unnecessary computational costs caused by excessively long vectors.
    \item \textbf{Weighted MLP Compression}: Based on the MLP Compression method, we perform weighted concatenation of the four vectors to obtain a 512D weighted vector.
    The weights are determined by the importance of each feature verified in the multi-feature fusion ablation experiments. 
\end{itemize}
As shown in Table 9, weighted fusion generally outperforms unweighted fusion, demonstrating that weighting effectively amplifies the discriminative power of more important features. Although employing MLPs for feature dimension unification can eliminate the impact of varying feature dimensions on results, the compression approach inevitably incurs information loss during feature reduction, which adversely affects the training of attack models. 

Additionally, we experimented with PCA for dimensionality reduction to transform high-dimensional features into lower-dimensional representations, but PCA dimensionality reduction damages the vector representation, leading to degraded attack performance. In contrast, upsampling the low-dimensional features followed by weighted concatenation significantly improves the attack effectiveness. We also try to train four attack models using single features separately, with the final inference decision generated through voting among these models. But single-feature attack models lose the complementary discriminative information from multi-feature interactions, leading to poor individual classifier performance and suboptimal voting outcomes.

\textbf{Data Illustration}. Based on the existing MovieLens and Book-Crossing datasets, we constructed natural language dataset formats for Last.FM and Delicious using the same prompt structure. Items were selected in chronological order of user history, with the most recent items used to determine user preferences and the remaining items serving as user history. Since Last.FM and Delicious only contain user historical records without explicit positive/negative item preferences (unlike MovieLens and Book-Crossing), the constructed data from Last.FM and Delicious contains less effective information than that from MovieLens and Book-Crossing. This may explain why in some experimental results, the data volume of these two datasets does not reach the expected attack effectiveness.
As shown in Figure 5, we illustrate data samples of the datasets we adopted from existing work (Book-Crossing and MovieLens) and our extended datasets (Last.FM and Delicious).

\section{Conclusion and Future Work}
In this paper, we propose a novel MIA paradigm against LLM-based recommendation systems. We obtain reference model using knowledge distillation to enhance the distinction between member and non-member data. We expand natural language datasets in recommendation field and generate stronger fused feature for attack model. Comprehensive experiments on various datasets demonstrate the superiority of our paradigm.

Currently our paradigm operates under gray-box background knowledge for MIA. Black-box distillation may provide a more rigorous attack premise. Our extended dataset only contains user interaction histories without user preferences, resulting in less semantically rich data. In future research, we will explore effective methods to extract higher-quality semantic information to enhance attack performance on semantically sparse datasets. We will also adapt our MIA paradigm to other LLM-based recommendation tasks to prove the scalability of our paradigm. Moreover, as discussed in the ``Experiments-Main Results'', the output distributions of models with different architectures on data features do not all conform to a Gaussian distribution. The reasons behind this phenomenon represent an important question for our future research.

\section{Acknowledgments}
This work was supported in part by the National Key Research and Development Program of China under Grant (2023YFC3310700), the National Natural Science Foundation of China (62572040, 62202041), the Beijing Natural Science Foundation (JQ24019), and the Fundamental Research Funds for the Central Universities (2025JBMC011).

\bibliography{aaai2026}

\end{document}